\documentclass{article}
\usepackage{amsmath, amssymb, amsfonts, amsthm}
\usepackage{bm}
\usepackage{geometry}
\usepackage{graphicx}
\usepackage{mathrsfs}

\usepackage{mathtools}
\usepackage{tikz}
\usetikzlibrary{decorations.pathmorphing, arrows.meta, positioning, calc, shadows, fit, backgrounds, 3d}
\usepackage{xcolor}
\usepackage[
  colorlinks=true,
  linkcolor=blue,
  citecolor=blue,
  urlcolor=cyan
]{hyperref}
\usepackage{cleveref}

\newcommand{\Vol}{\mathscr{V}}
\newcommand{\uvec}{\mathbf{u}}
\newcommand{\zhat}{\mathbf{\hat{z}}}

\newcommand{\Zvec}{\mathbf{Z}}
\newcommand{\omvec}{\boldsymbol{\omega}}
\newcommand{\omvect}{\boldsymbol{\omega}_t}

\newcommand{\norm}[1]{\left\lVert#1\right\rVert}

\title{A Variational Principle for the Vorticity Equation}
\author{A. Farooq}
\date{}

\begin{document}
\maketitle

\begin{abstract}
We present a variational formulation of the incompressible vorticity equation based on Gauss's principle of least constraint using  the Gauss constraint functional $\Zvec_\omega$. The central result is the Euler--Lagrange equation $\Zvec_\omega = -\nabla h$, where $h = \uvec\cdot\omvec$ is the helicity density. This reveals that the helicity gradient $\nabla h$ acts as the constraint force maintaining the solenoidality of the vorticity field, exactly as the pressure gradient $\nabla p$ maintains incompressibility in Taha et al.'s pressure-gradient minimization principle. The helicity density naturally emerges as the Lagrange multiplier enforcing $\nabla\cdot\omvec=0$, and at the solution the flow minimizes the norm of the helicity gradient $\|\nabla h\|^2$. This establishes the exact duality: helicity is to vorticity as pressure is to velocity. We apply the variational principle to the Burgers vortex and verify the the Euler-Lagrange equation. The variational principle connects to Moffatt's helicity conservation theorem, Arnold's geometric formulation of ideal fluid flow, and Kambe's gauge-theoretic formulation. This work provides a unified variational framework for fluid dynamics that spans classical mechanics, geometric mechanics, and topological field theory. We discuss how this work may provide a theoretical foundation for understanding the role of helicity gradients in boundary layer dynamics, with potential implications for the formation of coherent structures and the onset of transition. These applications are reserved for future work.
\end{abstract}

\section{Introduction}

Variational principles have played a central role in the development of fluid mechanics, offering deep insights into the governing equations and their underlying mathematical structure. Seliger and Whitham~\cite{Seliger1968} explored Clebsch representations for dissipative and non-self-adjoint systems and established connections between fluid variational principles and those of electrodynamics. Around the same time, Arnold~\cite{Arnold1966} delivered a landmark contribution by revealing that the Euler equations for an ideal incompressible fluid describe geodesic flow on the infinite-dimensional group of volume-preserving diffeomorphisms. This geometric insight transformed fluid dynamics into a branch of differential geometry, linking the motion of fluids to the curvature of configuration space. Earlier, Herivel~\cite{Herivel1955} had already pioneered variational approaches for dissipative systems, extending Hamilton's principle to derive the equations of motion for ideal fluids. 

In another thread Jackiw~\cite{Jackiw1990} brought a particle physics perspective and explored supersymmetry and non-Abelian generalizations and proposed  a Lagrangian involving Chern--Simons densities. Kambe~\cite{Kambe2007} developed a gauge-theoretic formulation of ideal fluid flows, treating rotational symmetries as gauge transformations and deriving the vorticity equation as a gauge field equation. 

Recently, Taha et al.~\cite{Taha2023} demonstrated that the incompressible Navier--Stokes equations can be derived from a variational principle in which the pressure $p$ acts as a Lagrange multiplier enforcing the incompressibility constraint $\nabla\cdot\uvec=0$. The resulting Gaussian functional minimized the norm of the pressure gradient.

In this work, we extend this idea to a variational formilation of the vorticity equation. The paper is organized as follows. In Section~2, we present the variational principle and then in Section 3,4 and 5 we examine its ramificaions. 

\section{A Variational Principle for the Vorticity Equation}

We propose a variational principle for the vorticity equation based on Gauss's principle of least constraint. The flow minimizes the squared norm of an unconstrained acceleration field, subject to the solenoidal constraint \(\nabla\cdot\omvec_t=0\).

A subtle but important point must be clarified regarding this constraint. While \(\nabla\cdot\omvec = \nabla\cdot(\nabla\times\uvec) \equiv 0\) is a strict kinematic identity for any continuous vector field, in the variational calculus we treat \(\omvec_t\) as an independent field to be varied. The space of admissible variations \(\delta\omvec_t\) may contain trial fields that momentarily violate solenoidality. The constraint thus acts to restrict these variations, ensuring the dynamics remain on the divergence-free manifold—precisely analogous to the role of the pressure in maintaining \(\nabla\cdot\uvec=0\).

Consider the unconstrained vector field:

\begin{equation}
\Zvec_\omega = \omvect - \nu \nabla^2 \omvec - 2(\omvec\cdot\nabla)\uvec - \uvec \times (\nabla \times \omvec)
\label{eq:Zom_def}
\end{equation}

which represents the rotational dynamics before the solenoidal constraint is imposed. We define the constraint Gaussian functional (see \cite{Lanczos2012}) as:

\[
\mathscr{G}_\omega = \frac{1}{2}\int_\Omega \norm{\Zvec_\omega}^2 \, d\Vol
\]

We introduce the Lagrangian

\begin{equation}
\mathscr{L}_\omega[\omvec, \omvec_t, \lambda] = \frac{1}{2} \int_\Omega |\Zvec_\omega|^2 \, d\Vol - \int_\Omega \lambda \, \nabla\cdot\omvec_t \, d\Vol
\label{eqn:lagrangian}
\end{equation}

where \(\lambda\) is a Lagrange multiplier enforcing \(\nabla\cdot\omvec_t=0\).

\subsection{Euler--Lagrange Equation}

We now derive the Euler--Lagrange equation from the Lagrangian (Eq.~\ref{eqn:lagrangian}).  Taking the variation with respect to the independent field \(\omvec_t\) gives:

\begin{equation}
\delta_{\omvect} \mathscr{L}_\omega[\omvec, \omvec_t, \lambda]
= \int_\Omega \Zvec_\omega \cdot \delta\omvec_t \, d\Vol - \int_\Omega \lambda \, \nabla\cdot(\delta\omvec_t) \, d\Vol
\end{equation}

since \(\Zvec_\omega\) depends linearly on \(\omvec_t\). Applying the product rule \(\nabla\cdot(\lambda \,\delta\omvec_t) = \nabla\lambda \cdot \delta\omvec_t + \lambda \nabla\cdot(\delta\omvec_t)\) allows us to rewrite the the second term in the variation as:

\begin{equation}
- \int_\Omega \lambda \, \nabla\cdot(\delta\omvec_t) \, d\Vol
= - \int_\Omega \nabla\cdot(\lambda \,\delta\omvec_t) \, d\Vol
+ \int_\Omega \nabla\lambda \cdot \delta\omvec_t \, d\Vol
\end{equation}

The first term on the right is a total divergence. Applying the divergence theorem converts it to a surface integral over the boundary \(\partial\Omega\):

\begin{equation}
\int_\Omega \nabla\cdot(\lambda \,\delta\omvec_t) \, d\Vol
= \oint_{\partial\Omega} \lambda \, \delta\omvec_t \cdot \mathbf{n} \, dS
\end{equation}

where \(\mathbf{n}\) is the outward unit normal. Assembling the full variation gives:

\begin{equation}
\delta_{\omvec_t}\mathscr{L}_\omega
= \int_\Omega (\Zvec_\omega + \nabla \lambda) \cdot \delta\omvec_t \, d\Vol
- \oint_{\partial\Omega} \lambda \, \delta\omvec_t \cdot \mathbf{n} \, dS
\end{equation}

The surface integral vanishes under standard boundary conditions: no-slip walls where the helicity vanishes on the boundary, fields decaying sufficiently fast at infinity, or periodic domains where the flux cancels. This is a standard requirement in variational calculus and defines the domain of validity of the principle. With the surface term eliminated, stationarity of the action requires the volume integral to vanish for all admissible variations \(\delta\omvec_t\). The fundamental lemma of the calculus of variations then yields the Euler--Lagrange equation pointwise:

\begin{equation}
\Zvec_\omega =- \nabla \lambda
\label{eq:EL_lambda}
\end{equation}

\subsection{Identification of the Lagrange Multiplier}

To identify \(\lambda\), we take the divergence of Eq.~\ref{eq:EL_lambda}:

\begin{equation}
\nabla^2 \lambda = 2\nabla\cdot((\omvec\cdot\nabla)\uvec) + \nabla\cdot(\uvec \times (\nabla \times \omvec))
\label{eq:Poisson_lambda}
\end{equation}

Now consider the helicity density \(h = \uvec\cdot\omvec\). Using the vector identity

\begin{equation}
\nabla h = (\uvec\cdot\nabla)\omvec + (\omvec\cdot\nabla)\uvec + \uvec\times(\nabla\times\omvec)
\end{equation}

and taking the divergence, we find

\begin{equation}
\nabla^2 h = \nabla\cdot((\uvec\cdot\nabla)\omvec) + \nabla\cdot((\omvec\cdot\nabla)\uvec) + \nabla\cdot(\uvec\times(\nabla\times\omvec))
\end{equation}

For an incompressible flow with solenoidal vorticity, the advective and stretching terms are equal:

\begin{equation}
\nabla\cdot((\omvec\cdot\nabla)\uvec) = \nabla\cdot((\uvec\cdot\nabla)\omvec)
\end{equation}

This follows directly from the Einstein summation convention: both expressions reduce to \((\partial_i u_j)(\partial_j \omega_i)\) after applying incompressibility and solenoidality. Hence,

\begin{equation}
\nabla^2 h = 2\nabla\cdot((\omvec\cdot\nabla)\uvec) + \nabla\cdot(\uvec\times(\nabla\times\omvec))
\end{equation}

Comparing with Eq.~\ref{eq:Poisson_lambda}, we have \(\nabla^2 \lambda = \nabla^2 h\). Thus \(\lambda = h + \phi\), where \(\nabla^2\phi = 0\). Under standard boundary conditions—no-slip walls where \(h=0\), or fields decaying at infinity—the harmonic function is uniquely determined and can be set to zero. In periodic domains, a constant harmonic component may persist, corresponding to a uniform helicity background, but this does not affect the dynamics since only \(\nabla h\) appears in the Euler--Lagrange equation. We therefore identify

\begin{equation}
\lambda = h = \uvec\cdot\omvec
\end{equation}

Substituting into Eq.~\ref{eq:EL_lambda} yields the central result:

\begin{equation}
\Zvec_\omega = -\nabla h
\label{eq:Zom_final}
\end{equation}

The flow minimizes the gradient of the helicity—a topological invariant—to maintain the solenoidal constraint.

\subsection{Recovery of the Vorticity Equation}

Substituting Eq.~\ref{eq:Zom_def} and the expanded gradient of helicity into Eq.~\ref{eq:Zom_final} gives
\begin{equation}
\omvec_t - \nu \nabla^2 \omvec - 2(\omvec\cdot\nabla)\uvec - \uvec \times (\nabla \times \omvec)
= -(\uvec\cdot\nabla)\omvec - (\omvec\cdot\nabla)\uvec - \uvec\times(\nabla\times\omvec)
\end{equation}

Cancelling the cross-product terms and rearranging yields the classical vorticity equation:
\begin{equation}
\omvec_t + (\uvec\cdot\nabla)\omvec - (\omvec\cdot\nabla)\uvec = \nu \nabla^2 \omvec
\end{equation}
The correct nonlinear advection and vortex stretching emerge naturally from the minimization of the helicity gradient.  The vorticity equation is thus the result of the minimization:

\[
\mathscr{G}_\omega = \frac{1}{2}\int_\Omega \norm{\nabla h}^2 \, d\Vol
\]

\subsection{Remark on the Sign Freedom in the Variational Principle}

The variational formulation admits a sign freedom. If we define the Gaussian as

\begin{equation}
\Zvec_\omega^\prime = \omvec_t - \nu \nabla^2 \omvec + 2(\uvec\cdot\nabla)\omvec + \uvec \times (\nabla \times \omvec)
\end{equation}

and the Lagrangian as

\begin{equation}
\mathscr{L} = \frac{1}{2}\int_\Omega \|\Zvec_\omega^\prime\|^2 \, d\Vol + \int_\Omega h \, \nabla\cdot\omvec_t \, d\Vol
\end{equation}

then variation with respect to $\omvec_t$ yields $\Zvec_\omega^\prime = \nabla h$. Using the vector identity, this reduces to the same vorticity equation.

This freedom reflects the fact that the helicity gradient $\nabla h$ acts as the constraint force, and only its magnitude $\|\nabla h\|^2$ is minimized at the solution. The sign is a matter of convention, analogous to the arbitrary sign of a Lagrange multiplier in constrained mechanics. This is also consistent with the fact that helicity $h = \uvec\cdot\omvec$ changes sign under parity transformations, confirming the robustness of the formulation.

\section{Relationship to Moffatt's Helicity Conservation Theorem}

Moffatt \cite{Moffatt1969} proved that the global helicity defined as:

\[
H = \int_\Omega \uvec\cdot\omvec \, d\Vol
\]

is conserved for inviscid, barotropic flows. 
This is a cornerstone result of topological fluid dynamics, linking the integral of the helicity density to the knottedness and linkage of vortex lines.

Our variational principle establishes a deeper, \emph{local} relationship to this classical theorem. 
The Euler-Lagrange equation $\Zvec_\omega = -\nabla h$ is equivalent to the full vorticity equation. 
Taking the dot product of the vorticity equation with $\uvec$ and integrating over the domain yields, in the inviscid limit,

\[
\frac{d}{dt} \int_\Omega \uvec\cdot\omvec \, d\Vol = 0
\]

To verify this, take the dot product of the vorticity equation with $\uvec$:

\[
\uvec\cdot\omvec_t + \uvec\cdot[(\uvec\cdot\nabla)\omvec - (\omvec\cdot\nabla)\uvec] = \nu \uvec\cdot\nabla^2\omvec
\]

Using the identity

\[
\uvec\cdot[(\uvec\cdot\nabla)\omvec - (\omvec\cdot\nabla)\uvec] = \nabla\cdot(\uvec h) - \uvec\cdot\nabla h - h\nabla\cdot\uvec
\]

and noting that $\nabla\cdot\uvec=0$ and $\uvec\cdot\nabla h = \nabla\cdot(h\uvec)$, the second term vanishes identically. Integrating over the domain and using the divergence theorem yields

\[
\frac{d}{dt}\int_\Omega h\,d\Vol = -\nu \int_\Omega \nabla\uvec:\nabla\omvec\,d\Vol
\]

In the inviscid limit ($\nu=0$), this gives Moffatt's theorem. Thus, global helicity conservation is a direct consequence of the minimization principle.

However, the converse is not true. 
Global helicity conservation is a single scalar constraint on the integral of the helicity density. 
It does not constrain the local distribution of $h$ or the magnitude of its gradient. 
A flow could conserve global helicity while having arbitrarily large local variations in $h$, and thus arbitrarily large $|\nabla h|$.

Our principle goes further: it states that the gradient of the helicity density is the constraint force that the flow locally minimizes. 
The flow is driven toward a state where $|\nabla h|$ is as small as possible, i.e., where the helicity density is as uniform as possible while respecting the topological constraints. 
This is a refinement of Moffatt's theorem: helicity conservation is not just a global statement; it is a \emph{local} dynamical principle that shapes the evolution of the flow.

In summary:
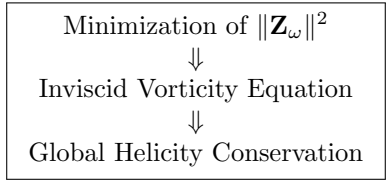
\[
\boxed{
\begin{array}{c}
\text{Minimization of } \|\Zvec_\omega\|^2 \\
\Downarrow \\
\text{Inviscid Vorticity Equation} \\
\Downarrow  \\
\text{Global Helicity Conservation}
\end{array}
}
\]
The implication does not reverse: global helicity conservation does not imply minimization of $|\nabla h|$. 
The variational principle provides the stronger, local statement that complements Moffatt's global theorem.

\section{Comparison to Taha's Variational Principle (PGMP) for the Navier--Stokes Equations}
\label{sec:comparison}

Taha et al.~\cite{Taha2023} demonstrated that the incompressible Navier--Stokes equations can be derived from a variational principle based on Gauss's principle of least constraint. For an incompressible, Newtonian fluid, the Navier--Stokes momentum equation is $
\rho\left(\uvec_t + \uvec\cdot\nabla\uvec\right) = \nabla\cdot\boldsymbol{\tau} - \nabla p, \qquad \nabla\cdot\uvec = 0,
$, where $\boldsymbol{\tau} = \mu(\nabla\uvec + (\nabla\uvec)^T)$ is the viscous stress tensor.  Following the structure of Gauss's principle,  the Gaussian is defined as the squared norm of the unconstrained acceleration:

\begin{equation}
\Zvec_u := \rho\left(\uvec_t + \uvec\cdot\nabla\uvec\right) - \nabla\cdot\boldsymbol{\tau}
\label{eq:Zu_def}
\end{equation}

This quantity represents the acceleration the fluid would have if there were no pressure constraint. The constraint force that maintains incompressibility is the pressure gradient $\nabla p$, which acts as the Lagrange multiplier.  The variational principle can be written as:

\begin{equation}
\mathscr{L}_u[\uvec, \uvec_t, p] = \frac{1}{2}\int_\Omega \norm{\Zvec_u}^2 \, d\Vol - \int_\Omega p \, \nabla\cdot\uvec_t \, d\Vol \label{eq:Lag_Taha}
\end{equation}

The second term enforces the incompressibility constraint $\nabla\cdot\uvec_t = 0$, with the pressure $p$ acting as the Lagrange multiplier.  This allows us to define the Gaussian which minimizes the constraint force

\[
\mathscr{G}_{u} = \frac{1}{2}\int_\Omega \norm{\nabla p}^2 \, d\Vol
\]

subject to $\nabla\cdot\uvec=0$. At the solution, $\Zvec_u = -\nabla p$, so the constraint force is the pressure gradient, and its norm is minimized.

We are now ready to see the duality between the two variational principles which are summarized in the table below. 

\begin{table}[htbp]
\centering
\begin{tabular}{l|c|c}
\hline
\textbf{Quantity} & \textbf{Primitive Formulation} & \textbf{Vorticity Formulation } \\
\hline
Constraint & $\nabla\cdot\uvec = 0$ & $\nabla\cdot\omvec = 0$ \\
Unconstrained dynamics & $\Zvec_u = \rho(\uvec_t + \uvec\cdot\nabla\uvec) - \nabla\cdot\boldsymbol{\tau}$ & $\Zvec_\omega = \omvect + (\uvec\cdot\nabla)\omvec - (\omvec\cdot\nabla)\uvec - \nu\nabla^2\omvec$ \\
Gaussian & $\mathscr{G}_{u} = \frac{1}{2}\int_\Omega \norm{\Zvec_u}^2 \, d\Vol$ & $\mathscr{G}_{\omega}=\frac{1}{2}\int_\Omega \norm{\Zvec_\omega}^2 \, d\Vol$ \\
Lagrange multiplier & Pressure $p$ & Helicity $h = \uvec\cdot\omvec$ \\
Constraint force & $\nabla p$ & $\nabla h$ \\
Euler--Lagrange equation & $\Zvec_u + \nabla p = 0$ & $\Zvec_\omega + \nabla h = 0$ \\
Resulting equation & $\Zvec_u = -\nabla p$ & $\Zvec_\omega = -\nabla h$ \\
Minimized  & $\norm{\nabla p}^2$ & $\norm{\nabla h}^2$ \\
\hline
\end{tabular}
\caption{The exact duality between the pressure-gradient minimization principle and the vorticity variational principle. Both formulations minimize the norm of the constraint force gradient.}
\label{tab:duality}
\end{table}

\section{Application to the Burgers Vortex}

Most of the well known solutions of Navier Stokes are 2D for which the helicity, $h=0$ making them unsuitable for testing the variational principle. The Burgers vortex is one of the few known 3D solutions of the Navier Stokes and we now demonstrate that it satisfies the variational principle $\Zvec_\omega = -\nabla h$. This provides a concrete, analytical verification of the principle.

\subsection{The Burgers Vortex Solution}

The Burgers vortex is a steady, axisymmetric solution of the Navier--Stokes equations, representing a stretched vortex in a background straining flow. In cylindrical coordinates $(r,\theta,z)$, the velocity field is:

\begin{align}
u_r &= -\frac{\alpha}{2} r    \\
u_\theta &= \frac{\Gamma}{2\pi r} \left(1 - e^{-\frac{\alpha r^2}{4\nu}}\right)   \\
u_z &= \alpha z 
\end{align}

where $\alpha > 0$ is the constant strain rate, $\Gamma$ is the circulation, and $\nu$ is the kinematic viscosity.

The vorticity field has only a $z$-component:

\begin{equation}
\omega_z = \frac{\Gamma \alpha}{4\pi \nu} e^{-\frac{\alpha r^2}{4\nu}} \equiv \omega_z(r)
\label{eq:om_z}
\end{equation}

The helicity density is:

\begin{equation}
h = \uvec \cdot \omvec = u_z \omega_z = \alpha z \, \omega_z(r)
\label{eq:helicity}
\end{equation}

This is non-zero and varies linearly with $z$.

\subsection{Computation of $\Zvec_\omega$}

Recall the definition of the Gaussian for the vorticity formulation (Eq.~\ref{eq:Zom_def}):
\begin{equation}
\Zvec_\omega = \omvec_t - \nu \nabla^2 \omvec - 2(\omvec\cdot\nabla)\uvec - \uvec\times(\nabla\times\omvec) \nonumber
\end{equation}

We compute each term for the Burgers vortex. For the first term, since the flow is steady, $\omvec_t = 0$.  For the second term,  $\omvec = \omega_z(r)\,\zhat$, we have:

\begin{equation}
\nabla^2 \omvec = \left( \frac{1}{r}\frac{d}{dr}\left(r \frac{d\omega_z}{dr}\right) \right) \zhat \nonumber
\end{equation}

From Eq.~\ref{eq:om_z}, $\omega_z(r) = C e^{-a r^2}$ with $C = \Gamma \alpha/(4\pi\nu)$ and $a = \alpha/(4\nu)$. Then:

\begin{equation}
\frac{d\omega_z}{dr} = -2a r \omega_z 
\label{eq:om_prime}
\end{equation}

and

\begin{equation}
\nabla^2 \omega_z = 4a (a r^2 - 1) \omega_z \nonumber
\end{equation}

Thus:

\begin{equation}
-\nu \nabla^2 \omvec = -\nu \nabla^2 \omega_z \, \zhat = -\nu \left(4a (a r^2 - 1) \omega_z\right) \zhat = \left(\alpha - \frac{\alpha^2 r^2}{4\nu}\right) \omega_z \,\zhat
\label{eq:term2}
\end{equation}

For the third non-linear term, since $\omvec = \omega_z \zhat$ and $\uvec$ has no $z$-dependence except $u_z = \alpha z$:

\begin{equation}
(\omvec\cdot\nabla)\uvec = \omega_z \frac{\partial \uvec}{\partial z} = \omega_z (0,0,\alpha) = \alpha \omega_z \,\zhat \nonumber
\end{equation}

Thus:

\begin{equation}
-2(\omvec\cdot\nabla)\uvec = -2\alpha \omega_z \,\zhat
\label{eq:term3}
\end{equation}

For the fourth term, first compute $\nabla\times\omvec$. Since $\omvec = (0,0,\omega_z(r))$:

\begin{equation}
\nabla\times\omvec = \left(0, -\frac{d\omega_z}{dr}, 0\right) = (0, -\omega_z', 0) \nonumber
\end{equation}

Now compute $\uvec\times(\nabla\times\omvec)$ using $\uvec = (u_r, u_\theta, u_z)$:

\begin{equation}
\uvec\times(0, -\omega_z', 0) = (u_z \omega_z', \, 0, \, -u_r \omega_z') \nonumber
\end{equation}

Substituting $u_z = \alpha z$ and $u_r = -\alpha r/2$:

\begin{equation}
\uvec\times(\nabla\times\omvec) = \left(\alpha z \omega_z', \, 0, \, \frac{\alpha r}{2} \omega_z' \right) \nonumber
\end{equation}

Using Eq.~\ref{eq:om_prime}, $\omega_z' = -2a r \omega_z = -\frac{\alpha}{2\nu} r \omega_z$. Then:

\begin{equation}
-\uvec\times(\nabla\times\omvec) = \left(-\alpha z \omega_z', \, 0, \, \frac{\alpha^2 r^2}{4\nu} \omega_z \right)
\label{eq:term4}
\end{equation}

Now sum Eq.~\ref{eq:term2},\ref{eq:term3},\ref{eq:term4}:

\begin{align}
\Zvec_\omega
&= \left[ \left(\alpha - \frac{\alpha^2 r^2}{4\nu}\right) \omega_z - 2\alpha \omega_z + \frac{\alpha^2 r^2}{4\nu} \omega_z \right] \zhat + (-\alpha z \omega_z') \hat{\mathbf{r}} \nonumber \\
&= -\alpha \omega_z \zhat - \alpha z \omega_z' \hat{\mathbf{r}} \nonumber
\end{align}

Thus:

\begin{equation}
\Zvec_\omega = -\alpha z \omega_z' \, \hat{\mathbf{r}} - \alpha \omega_z \, \zhat 
\label{eq:Z_om_final}
\end{equation}

\subsection{The Helicity Gradient}

From Eq.~\ref{eq:helicity}, $h = \alpha z \omega_z(r)$. Its gradient is:

\begin{equation}
\nabla h = \frac{\partial h}{\partial r} \hat{\mathbf{r}} + \frac{\partial h}{\partial z} \zhat = \alpha z \omega_z' \hat{\mathbf{r}} + \alpha \omega_z \zhat \nonumber
\end{equation}

Thus:

\begin{equation}
-\nabla h = -\alpha z \omega_z' \hat{\mathbf{r}} - \alpha \omega_z \zhat 
\label{eq:hel-gradient}
\end{equation}

Comparing Eq.~\ref{eq:Z_om_final} and Eq.~\ref{eq:hel-gradient}, we obtain:

\begin{equation}
\Zvec_\omega = -\nabla h \nonumber
\end{equation}

The Burgers vortex exactly satisfies the Euler--Lagrange equation $\Zvec_\omega = -\nabla h$. This provides a rigorous, analytical verification of the variational principle. The helicity gradient $\nabla h$ acts as the constraint force, and the Gaussian $\Zvec_\omega$ is precisely its negative. This confirms that the Burgers vortex is an extremum of the variational functional, and its helicity profile (a Gaussian in $r$, linear in $z$) is consistent with the minimization of the helicity gradient norm.

\section{Future Work}

The variational principle $\Zvec_\omega = -\nabla h$ suggests several promising avenues for future investigation.

First, the interpretation of the helicity gradient as the constraint force maintaining $\nabla\cdot\omvec=0$ raises the question of whether $\|\nabla h\|$ serves as a useful diagnostic for near-wall dynamics. At a solid wall, the no-slip condition forces $h=0$, while in the interior $h$ takes finite values, leading to a sharp helicity gradient in the boundary layer. This suggests that coherent structures such as streamwise vortices and streaks may be understood as mechanisms for transporting helicity and reducing this gradient. Direct numerical simulations (DNS) of turbulent boundary layers could test this hypothesis by computing the helicity gradient and its correlation with the generation and evolution of coherent structures.

Second, the variational principle may provide a new criterion for the onset of transition. If the helicity gradient $\|\nabla h\|$ exceeds a critical threshold, the flow may no longer be able to maintain a smooth helicity distribution, potentially triggering transition to turbulence. This hypothesis could be investigated through DNS of transitional boundary layers or channel flows.

Third, the framework suggests that drag reduction strategies—such as riblets, polymer additives, and superhydrophobic surfaces—may be effective because they reduce the helicity gradient near the wall. This interpretation could be tested by computing $\|\nabla h\|$ in DNS of drag-reducing flows.

Finally, the connection to the generalized isovorticity principle of Vladimirov, Moffatt, and Ilin \cite{Vladimirov1998} warrants further exploration. The minimization of $\|\nabla h\|^2$ may be the local, differential manifestation of the global energy minimization on isomagnetovortical folia. A rigorous investigation of this connection could provide a unified variational framework spanning both fluid dynamics and magnetohydrodynamics.

\section{Conclusion}

We have presented a variational formulation of the incompressible vorticity equation based on Gauss's principle of least constraint. The central result is the Euler--Lagrange equation

\[
\Zvec_\omega = -\nabla h
\]

where $\Zvec_\omega $ is the Gaussian—the unconstrained acceleration whose squared norm is minimized—and $h = \uvec\cdot\omvec$ is the helicity density. This equation reveals that the minimized residual is not zero, but rather the negative gradient of the helicity density. The helicity gradient $\nabla h$ thus acts as the constraint force that maintains the solenoidality of the vorticity field $\nabla\cdot\omvec=0$, exactly as the pressure gradient $\nabla p$ maintains incompressibility $\nabla\cdot\uvec=0$ in Taha's pressure-gradient minimization principle (PGMP).

Both principles share the same underlying structure: the Gaussian is the squared norm of the unconstrained dynamics—the evolution the system would follow in the absence of the constraint. The constraint is enforced by introducing a Lagrange multiplier (pressure or helicity). At the extremum, the unconstrained dynamics equal the negative gradient of the Lagrange multiplier, and the norm of this gradient is minimized.

The result $\Zvec_\omega = -\nabla h$ is interpreted as the statement that the vorticity dynamics evolve as constrained motion on an iso-helicity surface in the infinite-dimensional phase space of solenoidal vector fields. This is the fluid-dynamic analog of a particle moving on a constant-energy surface, where the constraint force is the gradient of the conserved quantity. The topological nature of helicity—as a measure of the knottedness and linkage of vortex lines—elevates this principle from a purely mechanical statement to one with deep geometric and topological content.

The variational principle provides a direct Eulerian connection to Arnold's geometric formulation of ideal fluid flow. In Arnold's framework, the Euler equations are geodesic flow on the group of volume-preserving diffeomorphisms, with helicity emerging as a Casimir invariant of the non-canonical Poisson bracket \cite{Morrison1998}. Our formulation reveals that the helicity gradient $\nabla h$ is the local constraint force that maintains the solenoidality of the vorticity field—the Eulerian manifestation of these topological invariants. The principle also connects naturally to Kelvin's circulation theorem, which emerges from the particle relabeling symmetry (see \cite{MarsdenRatiu1994}) encoded in the solenoidal constraint, and to Kambe's \cite{Kambe1985} gauge formulation, where helicity itself acts as the gauge field without the need for a vector potential.

In summary, this work establishes a unified variational framework for fluid dynamics that spans classical mechanics, geometric fluid dynamics, and topological field theory. By revealing that helicity is to vorticity as pressure is to velocity, it completes the duality between primitive-variable and vorticity formulations, providing a variational foundation for the topological invariants at the heart of fluid motion.

\end{document}